\documentclass[conference]{IEEEtran}
\IEEEoverridecommandlockouts
\usepackage{cite}
\usepackage{amsmath,amssymb,amsfonts}
\usepackage{algorithmic}
\usepackage{graphicx}
\usepackage{textcomp}
\usepackage{xcolor}
\usepackage{braket}
\usepackage{multirow}
\usepackage{subfig}
\usepackage{enumitem}
\usepackage{pgfplots}
\usepackage{todonotes}
\def\BibTeX{{\rm B\kern-.05em{\sc i\kern-.025em b}\kern-.08em
    T\kern-.1667em\lower.7ex\hbox{E}\kern-.125emX}}

\begin{document}
%\title{An Efficient Device-Independent Quantum Secure Direct Communication with User Authentication Protocol
\title{User-Authenticated Device-Independent Quantum Secure Direct Communication Protocol}

% \author{\IEEEauthorblockN{\Large{Nayana Das*, Saikat Basu*, Goutam Paul$\dag$, Vijay S Rao*}\\}
% \IEEEauthorblockA{\large{*LTIMindtree Ltd., Kolkata, India \\
% \{nayana1.das, saikat.basu, vijay.rao\}@ltimindtree.com\\
% $\dag$Cryptology and Security Research Unit, Indian Statistical Institute, Kolkata, India \\
% goutam.paul@isical.ac.in}}}

\author{\IEEEauthorblockN{Nayana Das, Saikat Basu}
\IEEEauthorblockA{Global Technology Office \\ 
\textit{LTIMindtree Ltd.}\\
Kolkata, India \\
\{nayana1.das,saikat.basu\}@ltimindtree.com}
\and
\IEEEauthorblockN{Goutam Paul}
\IEEEauthorblockA{Cryptology and Security Research Unit\\ 
\textit{Indian Statistical Institute}\\
Kolkata, India\\
goutam.paul@isical.ac.in}
\and
\IEEEauthorblockN{Vijay S. Rao}
\IEEEauthorblockA{Global Technology Office \\
\textit{LTIMindtree Ltd.}\\
Amsterdam, The Netherlands\\
vijay.rao@ltimindtree.com}
}

\maketitle

\begin{abstract}
Device-Independent Quantum Secure Direct Communication (DI-QSDC) enhances quantum cryptography by enabling secure message transmission without relying on the trustworthiness of the devices involved. This approach mitigates risks associated with compromised or untrusted devices, common in traditional quantum communication.

In this paper, we propose the first of its kind DI-QSDC protocol with user identity authentication. This ensures the authenticity of both the sender and receiver prior to message exchange. We then discuss the security of the proposed protocol against common attacks, demonstrating that no eavesdropper gains any information from either the quantum or the classical channel. Next, we implement the protocol on IBM's quantum hardware and evaluate its performance in a realistic noisy environment. Additionally, by simulating common attack models, we showcase that the protocol is secure against any eavesdropper in the channel. These findings highlight the protocol's robust security and practical feasibility for real-world secure quantum communication.

%Device-Independent Quantum Secure Direct Communication with User Authentication (UA-DI-QSDC) enhances quantum cryptography by enabling secure message transmission after verifying the identities of the communicating parties, without depending on the trustworthiness of the devices involved. This approach mitigates risks associated with compromised or untrusted devices, common in traditional quantum communication. While a recent paper proposed a first-of-its-kind UA-DI-QSDC protocol, the work did not conduct any security analysis.

%In this paper, We analyze the security of the UA-DI-QSDC protocol and show there is a security loophole in the protocol, which impacts the user authentication process. Accordingly, we propose an improved version of this protocol, which resolves the issue. 
% we present the first comprehensive security analysis of the UA-DI-QSDC protocol. First, we propose a simplified and efficient version of the existing protocol that requires fewer resources to implement, particularly single-qubit gates. 
%We then discuss the security of the proposed protocol against common attacks, demonstrating that no eavesdropper gains any information from either the quantum or the classical channel. Next, we implement the protocol on IBM's quantum hardware and evaluate its performance in a realistic noisy environment. Additionally, by simulating common attack models, we showcase that the protocol is secure against any eavesdropper in the channel. These findings highlight the protocol's robust security and practical feasibility for real-world quantum communication.

\end{abstract}


\section{Introduction}
Traditional communication protocols require a key exchange phase to secure the channel before message transmission. In contrast, quantum secure direct communication (QSDC) protocols~\cite{deng2003two,das2021quantum,das2022measurement} enable secure message transmission, encoded in quantum states,  without requiring any secret key. This method simplifies the design of secure communication protocols as it reduces potential vulnerabilities associated with key distribution and key management.

While traditional quantum communication protocols offer strong security, they often rely on the assumption that the devices used are perfect~\cite{xu2020secure}. In practical scenarios, this assumption is unrealistic due to device imperfections and potential side-channel attacks that can compromise communication security~\cite{gisin2002quantum,portmann2022security}. Device-independent (DI) quantum communication protocols address this issue by not relying on the trustworthiness of the quantum devices~\cite{acin2007device,zhen2023device}. Even if an eavesdropper controls the devices, DI protocols remain secure by leveraging the violation of Bell inequalities~\cite{bell1964einstein} and the non-local correlations of entangled quantum states. DI-QSDC can relax the security assumptions about the devices’ internal working, and effectively enhance QSDC’s security. The first DI-QSDC protocol was proposed by Zhou et al. in 2020~\cite{zhou2020device}, followed by several variants, such as hyperentanglement-based one-step DI-QSDC protocol~\cite{zhou2022one}, high-capacity DI-QSDC protocol based on hyper-encoding technique~\cite{zeng2023high}, and DI-QSDC protocol utilizing practical, highly efficient single-photon sources~\cite{zhou2023device}. %~\cite{zhou2020device,zhou2022one,zeng2023high,zhou2023device}.

%Despite these significant advances, no state-of-the-art DI-QSDC protocol addresses identity authentication, a critical component of secure communication as it prevents an eavesdropper from impersonating a legitimate party. 
Identity authentication is a critical component of secure communication as it prevents an eavesdropper from impersonating a legitimate party. The concept of quantum-based identity authentication was first introduced by Cr{\'e}peau et al.~\cite{crepeau1995quantum} in 1995, which utilizes quantum oblivious transfer~\cite{bennett1991practical}. The idea of integrating user authentication with QSDC was first proposed by Lee et al. in 2006, using Greenberger-Horne-Zeilinger (GHZ) states~\cite{lee2006quantum}. However, Zhang et al. later identified vulnerabilities in this protocol, particularly its susceptibility to intercept-and-resend attacks, and subsequently proposed a revised version to enhance its security~\cite{zhang2007comment}. Since then, several new QSDC protocols incorporating user authentication have been developed, further advancing the field~\cite{das2022cryptanalysis,das2021quantum,das2022measurement}.

%However, DI-QSDC protocols do not inherently support the authentication of the communicating parties and may lead to unauthorized access and impersonation attacks. To this end, 
Despite the above-mentioned advances in QSDC protocols, none of the protocols provide both device independence and identity authentication. In this paper, we introduce a novel user-authenticated DI-QSDC (UA-DI-QSDC) protocol, which to the best of our knowledge is the first of its kind to incorporate user authentication directly into the DI-QSDC framework, thereby enhancing the overall security of DI-QSDC. To enable user authentication, the sender and the receiver verify their pre-shared secret identities before the message transmission. Additionally, we conduct a comprehensive security analysis to ensure that our protocol effectively mitigates potential threats and vulnerabilities, making it a robust solution for secure quantum communication.

While most quantum communication protocols, including UA-DI-QSDC, assume closed quantum systems (i.e., isolated from the environment and noise-free), real quantum systems are open and susceptible to external noise. 
%Specifically, when we consider the current generation of quantum computers, that are extremely error-prone, called noisy intermediate scale quantum (NISQ)~\cite{Preskill2018quantumcomputingin} hardware. 
This is certainly the case with the current generation of quantum computers that are extremely error-prone, called noisy intermediate scale quantum (NISQ)~\cite{Preskill2018quantumcomputingin} hardware. 
These NISQ devices are limited by their qubit counts as well as short coherence time. Thus, execution of the protocol on NISQ devices gets impacted by the errors, particularly occurring due to decoherence, imperfect gate operations, state preparation, and measurements. We implemented our UA-DI-QSDC protocol in IBM's quantum hardware and studied its behavior in the presence of noise. Further, we have checked the security of the protocol by simulating some common attacks by an eavesdropper in the channel.

\textbf{Contributions.} Following our contributions in this paper:

$\bullet$ We enable user authentication (UA) on DI-QSDC to propose the first-ever UA-DI-QSDC protocol, which protects against impersonation attacks.

$\bullet$ Further, we analyzes the security of our proposed UA-DI-QSDC protocol and rigorously evaluate its resilience against common attacks.
%The first-of-its-kind UA-DI-QSDC protocol is proposed along with its comprehensive security analysis.

$\bullet$ To validate the practical applicability of our proposed UA-DI-QSDC protocol, we have emulated it on IBM's quantum hardware. It allows us to assess the protocol's performance in a real-world noisy quantum computing environment. We conduct a detailed performance analysis of the protocol under varying channel lengths providing us valuable insights into the protocol's efficiency and operational feasibility in practical conditions. 

Table~\ref{Comparison} compares our proposed UA-DI-QSDC protocol with the existing DI-QSDC works against key features. As shown, our protocol provides UA while just consuming just 1 qubit/message bit. 
\begin{table}[t]
\resizebox{\columnwidth}{!}{%
\begin{tabular}{|c|c|c|c|c|}
\hline
\textbf{Protocol}                                       & \textbf{\begin{tabular}[c]{@{}c@{}}Resource\\ type\end{tabular}} & \textbf{\begin{tabular}[c]{@{}c@{}}Measurement\\ for decoding\end{tabular}} & \textbf{\begin{tabular}[c]{@{}c@{}}No. of qubits\\ per message bit\end{tabular}} & \textbf{UA} \\ \hline
Zhou et al.~\cite{zhou2020device} & Entanglement                                                     & BSM                                                                         & $1$                                                                              & No          \\ \hline
Zhou et al.~\cite{zhou2022one}    & \begin{tabular}[c]{@{}c@{}}Hyper-\\ entanglement\end{tabular}    & BSM                                                                         & $1$                                                                              & No          \\ \hline
Zhou et al.~\cite{zhou2023device} & Single qubits                                                    & BSM                                                                         & $2$                                                                              & No          \\ \hline
Zeng et al.~\cite{zeng2023high}   & \begin{tabular}[c]{@{}c@{}}Hyper-\\ entanglement\end{tabular}    & HBSM                                                                        & $\frac{1}{2}$                                                                    & No          \\ \hline
Proposed protocol                                        & Entanglement                                                     & BSM                                                                         & $1$                                                                              & Yes         \\ \hline
\end{tabular}
}

\scriptsize{\vspace{1mm}*BSM: Bell state measurement, HBSM: Hyper-entanglement Bell state measurement}
\caption{Comparison between state-of-the-art DI-QSDC protocols and our proposed protocol}\label{Comparison}
\end{table}

%The proposed protocol is emulated in IBM's quantum hardware and a detailed performance analysis have been carried out. 

The rest of this paper is organized as follows: In Section~\ref{sec2}, we propose our novel UA-DI-QSDC protocol. In Section~\ref{security}, the security of the protocol is analyzed against all common attacks. Next, we emulate the proposed UA-DI-QSDC protocol and provide a comprehensive analysis of the performance of the protocol in Section~\ref{Sec: Implementation}. Finally, Section~\ref{conclusion} presents our concluding remarks and the future scope.

\begin{figure*}
    \centering
    \includegraphics[width=0.8\textwidth]{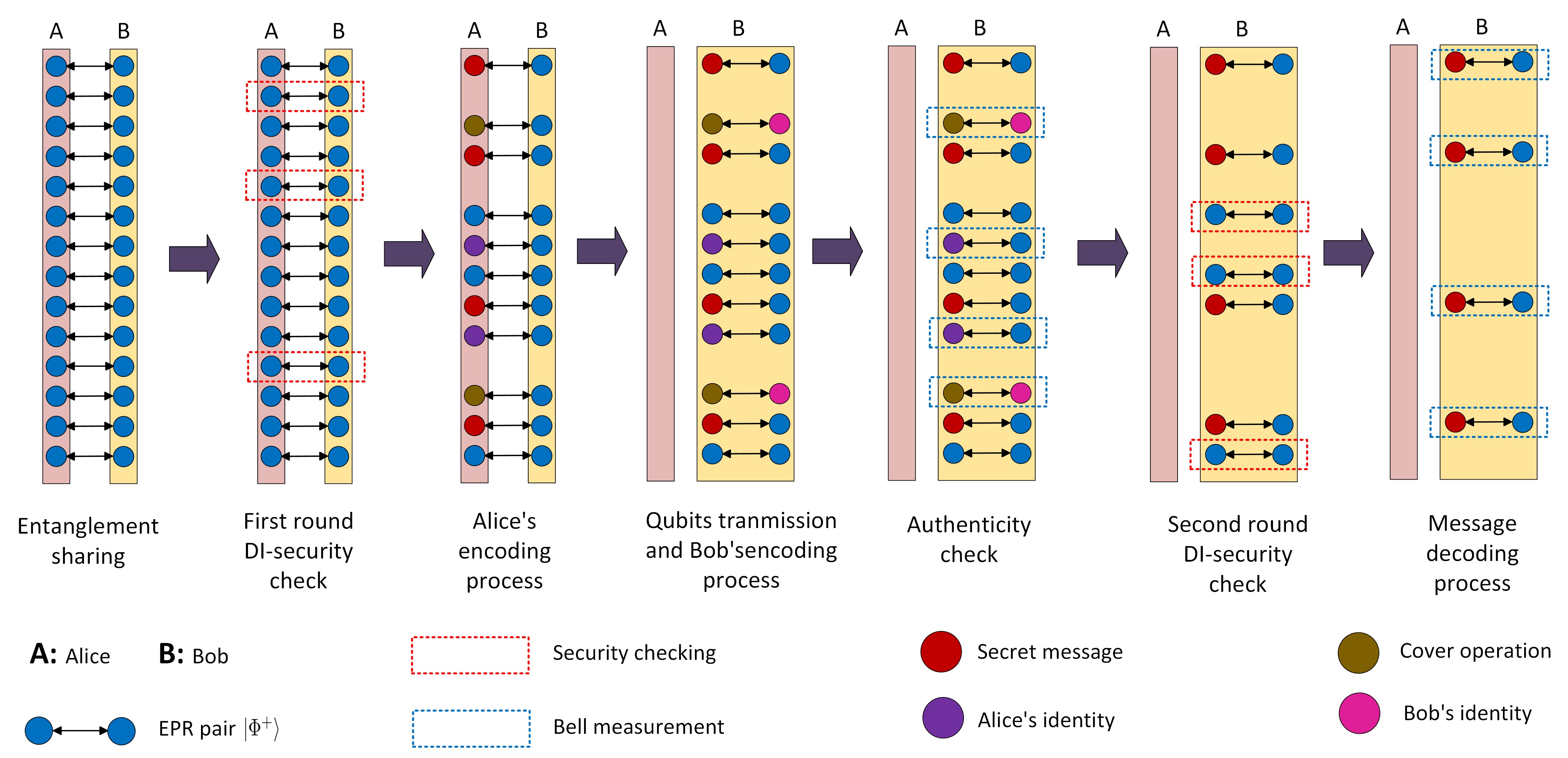}
    \caption{Schematic diagram of the proposed UA-DI-QSDC protocol}
    \label{fig: protocol}
\end{figure*}

\section{Proposed UA-DI-QSDC protocol}\label{sec2}
In the UA-DI-QSDC protocol, we consider two legitimate users, Alice (the sender) and Bob (the receiver). They are connected through a quantum and an authenticated classical channel for the transmission of quantum and classical information respectively. For identity authentication on the quantum channel, Alice and Bob have their pre-shared secret identities $\text{id}_A$ and $\text{id}_B$ ($2l$-bits each) respectively, which are exclusive to them.

Suppose Alice has an $n$-bit message $m$ that she wants to send to Bob through a quantum channel securely. Alice incorporates $c$ check bits into $m$ at random positions, forming a new bit string $m'$ with a length of $n + c=2N$, for some integer $N$. The other symbols and their descriptions used are provided in Table~\ref{symbols}. Fig.~\ref{fig: protocol} shows the schematic of the protocol, described as follows: 

\begin{table}[t]
\centering
\normalsize
\begin{tabular}{c|l}
\hline
\textbf{Notation}                                                                                        & \textbf{Description}                        \\ 
\hline
%$I$                                                                                                      & $\lvert 0 \rangle \langle 0 \rvert +\lvert 1 \rangle \langle 1 \rvert$             \\ 
% \hline
%$\sigma_x$                                                                                               & $\lvert 0 \rangle \langle 1 \rvert +\lvert 1 \rangle \langle 0 \rvert$             \\ 
% \hline
%$i\sigma_y$                                                                                              & $\lvert 0 \rangle \langle 1 \rvert -\lvert 1 \rangle \langle 0 \rvert$                     \\ 
% \hline
%$\sigma_z$                                                                                               & $\lvert 0 \rangle \langle 0 \rvert - \lvert 1 \rangle \langle 1 \rvert$                     \\ 
% \hline
$I, \sigma_x, \sigma_y, \sigma_z$                                                                        & Pauli unitary operators~\cite{nielsen00}                    \\ 
% \hline
$\ket{\Phi^{\pm}}$                                                                                       & $\frac{1}{\sqrt{2}}(\ket{00} \pm \ket{11})$ \\ 
% \hline
$\ket{\Psi^{\pm}}$                                                                                       & $\frac{1}{\sqrt{2}}(\ket{01} \pm \ket{10})$ \\ 
% \hline
$\ket{\Phi^{\pm}}, \ket{\Psi^{\pm}}$                                                                     & Bell states or EPR pairs                    \\ 
% \hline
$\{\ket{\Phi^{\pm}},\ket{\Psi^{\pm}}\}$                                                                  & Bell basis                                  \\
\multirow{2}{*}{$S^i_{CHSH}$} & CHSH polynomial in $i$-th                  \\                         & round DI-security check                                 \\
\hline
\end{tabular}

\caption{List of symbols}
\label{symbols}
\end{table}

\begin{enumerate}[wide, labelwidth=!, labelindent=0pt]   
    \item \textbf{Entanglement sharing:} Alice and Bob share a quantum channel with a source that emits EPR pairs in the $\ket{\Phi^+}$ state. They share $(N+2l+2d)$ EPR pairs, where $d$ EPR pairs are used for each round of the DI-security check process (several hundred to a few thousand pairs are needed to achieve a statistically significant result for a rough estimation of the CHSH value). Alice has a sequence $\mathbf{S}_A$ consisting of the first qubits of all $\ket{\Phi^+}$ states, and Bob has a sequence $\mathbf{S}_B$ containing all the partner qubits corresponding to $\mathbf{S}_A$.

    \item \textbf{First round of DI-security check:} Alice first randomly selects a sufficiently large subset of $d$ qubits from the sequence  $\mathbf{S}_A$ as the first round security checking qubits, and publishes their positions to Bob through a public classical channel, so that he can choose the corresponding partner qubits from $\mathbf{S}_B$ for security check. The remaining qubits are stored in a quantum memory device. Alice and Bob independently perform randomly chosen measurements on the security-checking qubits. 
    
    Alice has three possible measurement bases $\mathcal{B}_{A_j}=\ket{0} \pm e^{iA_j}\ket{1}$, for $j \in \{0, 1, 2\}$ with $A_0 =\frac{\pi}{4}, A_1 = 0, A_2 =\frac{\pi}{2}$ ($i=\sqrt{-1}$). Bob has two possible measurement bases $\mathcal{B}_{B_k}= \ket{0} \pm e^{iB_k}\ket{1}$, for $k \in \{1, 2\}$, with $B_1 =\frac{\pi}{4}, B_2 =-\frac{\pi}{4}$. All the measurement results $a_0, a_1, a_2, b_1, b_2$ have binary outcomes labeled by $\pm 1$. 
    
    Alice and Bob reveal their measurement bases and measurement results via the classical channel to estimate the CHSH polynomial~\cite{clauser1969proposed}, as given by, $$S^1_{CHSH}=\langle a_1 b_1 \rangle + \langle a_1 b_2 \rangle + \langle a_2 b_1 \rangle - \langle a_2 b_2 \rangle,$$ where $\langle a_j b_k \rangle=\Pr(a=b|jk) - \Pr(a \neq b|jk)$. Without the loss of generality, we suppose that the marginals are random for each measurement, i.e., $\langle a_j \rangle= \langle b_k \rangle=0$ $\forall j,k$. Ideally, $S^1_{CHSH}=2\sqrt{2}$, but due to channel noise, the value may slightly deviate, in that case, if $S^1_{CHSH}=2\sqrt{2} - \epsilon_1 > 2$ (where $0<\epsilon_1 < 2(\sqrt{2}-1)$ is the error parameter lower than the threshold value), then they continue the protocol, else they abort it.

   \item \textbf{Alice's encoding process:} Alice retrieves the stored photons from the memory device. Then, she randomly selects $d$ qubits for the second round of the DI-security check process and does not perform any operation on them. From the remaining $(N+2l)$ qubits, she randomly chooses $N$ qubits to encode the secret message, denoted as the set $M_A$, and $l$ qubits, denoted as the set $C_A$, to encode her secret identity $\text{id}_A$. The remaining $l$ qubits, denoted as the set $D_A$ are reserved for Bob to encode his identity $\text{id}_B$. 
   
   The encoding process for both the message $m'$ and the identity $\text{id}_A$ is the same. By applying a unitary operator, Alice encodes two bits of classical information into one qubit. To encode $00$, $01$, $10$ and $11$, she applies the Pauli operators~\cite{nielsen00} $I$, $\sigma_z$, $\sigma_x$ and $i\sigma_y$ respectively on her qubit. Additionally, Alice randomly applies cover operations from the set $\{I, \sigma_z, \sigma_x,i\sigma_y\}$ on the qubits of $D_A$. 

   \item \textbf{Authentication process:} 
   Alice successively sends all the qubits of $\mathbf{S}_A$ to Bob. Once Bob receives the sequence, Alice reveals the positions of the qubits in $D_A$ through the classical channel. Bob then encodes his identity $\text{id}_B$ on the corresponding partner qubits (say, $D_B$) by applying the same encoding rules that Alice used. After encoding, Bob performs Bell-state analysis~\cite{kwiat1998embedded} on the qubit pairs of $(D_A, D_B)$ and announces the results.

   Alice previously applied cover operations on the qubits in $D_A$, which transforms the initial $\ket{\Phi^+}$ state into one of the other Bell states. Consequently, when Bob announces the measurement results of $(D_A, D_B)$, it appears as a random Bell state, making $\text{id}_B$ reusable. Since Alice knows both $\text{id}_B$ and the applied cover operations, she can verify Bob's identity. If there is a significant error, Alice will abort the protocol, and Bob will not receive any secret message because he does not know the positions of the qubits in $\mathbf{S}_A$ that correspond to the secret message $m$.

   If the error rate is acceptable, Alice will disclose the positions of the qubits in $C_A$ that correspond to her identity $\text{id}_A$. Bob then performs Bell-state analysis on these qubits with their corresponding partner qubits in $\mathbf{S}_B$ to verify Alice's identity. If Bob detects a significant error, he will terminate the protocol. It is important to note that the measurement results corresponding to $\text{id}_A$ are not publicly disclosed, ensuring Alice's identity remains reusable.
  
   \item \textbf{Second round of DI-security check:} Alice publicly shares the positions of the security-checking qubits via the classical channel. Bob then extracts the corresponding partner qubits from $\mathbf{S}_B$ and performs the second round of DI-security check on his own. Specifically, Bob independently performs measurements on the two photons in each checking Bell pair, randomly choosing the bases $\mathcal{B}_{A_j}$ and $\mathcal{B}_{B_k}$ for his measurements. After the measurements, he can estimate the CHSH polynomial ($S^2_{CHSH}$). Similar to the first round, we also require $S^2_{CHSH}=2\sqrt{2} - \epsilon_2 > 2$ (where $0<\epsilon_2 < 2(\sqrt{2}-1)$ is the error parameter lower than the threshold value). If this condition is not met, Alice and Bob will discard the entire communication.
   
   \item \textbf{Message decoding process:} Bob discards all the measured qubits and performs Bell-state analysis on the remaining qubit pairs to decode the classical bit string $m'$. Alice and Bob publicly verify the random check bits to ensure the integrity of the messages. If the error rate is within acceptable limits, Bob receives the secret message $m$ and the communication process is completed. 
\end{enumerate}

\section{Security Analysis}\label{security}
In the device-independent scenario, we only require Eve to obey the laws of quantum physics, and no other limitations are imposed on her. Consequently, we assume that Eve controls the entanglement source and fabricates Alice’s and Bob’s measurement devices. In both the security checking rounds, Alice and Bob can only use the observed relation between the measurement basis selection (input) and outcomes to bound Eve’s knowledge. 

It is only required to analyzes the security of the protocol from Alice's encoding process. This is due to the first round of security check occurring after sharing the entangled qubits, which follows existing works~\cite{ekert1991quantum,acin2006efficient}. The security of UA-DI-QSDC is analyzed against five common attack strategies: impersonation, intercept-and-resend, entangle-and-measure, man-in-the-middle and information leakage attacks.

\subsection{Impersonation attack} 
In this attack model, an eavesdropper (Eve) is impersonating a legitimate party.

First, assume Eve impersonates Alice to send a message to Bob. Since Eve does not know the pre-shared key $\text{id}_A$, she  applies Pauli operators randomly on the qubits of $C_A$ instead of performing the correct unitary operation to encode $\text{id}_A$. When Bob receives, measures these qubits with their partner qubits from $C_B$ using Bell basis, Bob detects Eve's interference as he knows $\text{id}_{A}$. The probability of Bob detecting Eve is $1-(\frac{1}{4})^l$ as the chance of Eve applying the right unitary is $\frac{1}{4}$ for each of the $l$ qubits. 

Conversely, if Eve tries to impersonate Bob to receive the secret message from Alice, she faces a similar challenge. Without knowing $\text{id}_B$, Eve applies Pauli operators randomly on the qubits of $D_B$, correctly guessing the unitary with a probability of $\frac{1}{4}$ for each qubit. Eve then measures the qubits of $(D_A, D_B)$ in the Bell basis and announces the result. Alice, knowing the value of $\text{id}_B$, compares the measurement results with the expected results and detects Eve with a probability of $1 - (\frac{1}{4})^l$.

In both scenarios, the legitimate party (either Alice or Bob) can detect Eve's eavesdropping with probability $\rightarrow 1$ as $l~\rightarrow~\infty$ due to Eve's random application of Pauli operators and the legitimate parties' knowledge of the pre-shared keys.

\subsection{Intercept-and-resend attack}
In this attack model, Eve intercepts the qubits $\mathbf{S}_A$ sent from Alice to Bob through the quantum channel. She measures these intercepted qubits in the $\{\ket{u}, \ket{v} \}$ basis, where $\ket{u}$ and $\ket{v}$ are some orthogonal states. After measuring, she resends the qubits to Bob.

For any basis $\{\ket{u}, \ket{v} \}$, the Bell state $\ket{\Phi^+}$ can be expressed as:
$$\ket{\Phi^+} = \frac{1}{\sqrt{2}} (\ket{uu} + \ket{vv}).$$
Measuring the first qubit in the $\{\ket{u}, \ket{v} \}$ basis collapses the second qubit to either the $\ket{u}$ or $\ket{v}$ state, depending on the measurement outcome.
As a result, the joint state of the qubits becomes either $\ket{uu}$ or $\ket{vv}$, both of which are separable states. In a separable state, qubits do not exhibit quantum correlations that are characteristic of entanglement.

During the security check process, Alice and Bob test the quantum correlations by calculating the value of the CHSH polynomial. For entangled states, this value should exceed 2, but for separable states, it will not. Thus, when Eve's attack causes the states to be separable, the value of $S_{CHSH}^2$ calculated by Bob will not exceed $2$. Detecting this, Alice and Bob will recognize the presence of an eavesdropper and abort the protocol to ensure the security of their communication.

\subsection{Man-in-the-middle attack} 
In this attack strategy, Eve intercepts the sequence $\mathbf{S}_A$ from the quantum channel and keeps it. She then creates a new sequence $Q_E$ of single qubit states and sends this sequence to Bob instead of the original sequence $\mathbf{S}_A$. Because the qubits in $Q_E$ are not correlated with the qubits in $\mathbf{S}_B$ that Bob has, the calculated value of the CHSH polynomial will be $\leq 2$, indicating classical correlations rather than quantum entanglement. As a result, Alice and Bob can easily detect the presence of Eve during the security check process by observing this deviation from the expected quantum correlations.

\subsection{Entangle-and-measure attack}
As Alice and Bob share Bell states, Eve cannot successfully apply an entanglement-and-measure attack due to the monogamy of entanglement~\cite{coffman2000distributed}. This principle ensures that if two particles are maximally entangled, they cannot share significant entanglement with a third particle. Therefore, any attempt by Eve to entangle her ancillary qubit with Alice's or Bob's qubit will disturb the original entanglement. By estimating the CHSH polynomial, Alice and Bob can detect these disturbances as the value will drop below the expected threshold. Therefore, the monogamy of entanglement ensures that Eve's attack introduces detectable disturbances, proving the security of the communication.

\subsection{Information leakage attack} This refers to the information about the secret message that Eve might gain by intercepting classical channels. However, in the present protocol, the measurement outcomes associated with the secret bits are not transmitted over the classical channel. Consequently, Eve is unable to obtain any confidential information from the classical channels.

\section{Emulation of UA-DI-QSDC on Quantum hardware}\label{Sec: Implementation}
In the proposed protocol, multiple EPR pairs are shared between Alice and Bob. After verifying the channel's security, Alice encodes her message and identity into her qubits using Pauli unitary operators and sends these encoded qubits to Bob through a quantum channel ($U_{C}$). Once Bob receives the qubits, Alice and Bob authenticate each other and verify the channel's security. Bob then measures each of the qubits in the message-encoded EPR pairs in the Bell basis to retrieve the secret message from Alice. The entire execution of the protocol assumes $U_{C}\equiv I$, where $I$ is the Identity operation. However, in reality, the channel is noisy, represented as \(U_{C}^{n}\neq I\), which can affect Bob's measurement outcomes. Additionally, the NISQ devices may also experience state preparation and measurement errors.

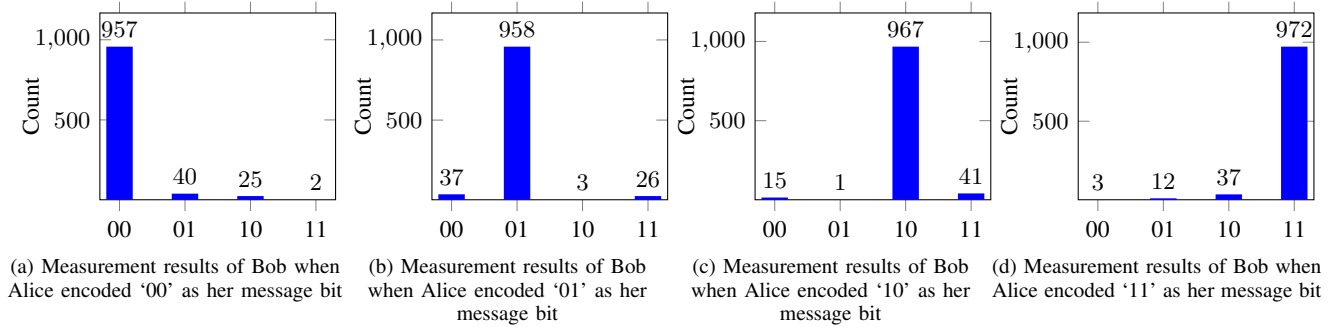
\begin{figure*}[t]
    \centering
    \subfloat[\centering Measurement results of Bob when Alice encoded `00' as her message bit]{
    \begin{tikzpicture}
        \tikzstyle{every node}=[font=\small]
        \begin{axis}[ybar,bar width = 10pt, width=0.26\textwidth, xtick={1,2,3,4},xticklabels={00,01,10,11},nodes near coords, enlarge y limits={upper, value=0.22}, ylabel={Count},y label style={at={(axis description cs:0.1,.5)}}]
            \addplot [draw=none,fill=blue] coordinates {
            (1, 957)
            (2, 40)
            (3, 25)
            (4, 2)
            };
        \end{axis}
    \end{tikzpicture}
    }%
    \subfloat[\centering  Measurement results of Bob when Alice encoded `01' as her message bit]{\begin{tikzpicture}
        \tikzstyle{every node}=[font=\small]
        \begin{axis}[ybar,bar width = 10pt, width=0.26\textwidth, xtick={1,2,3,4},xticklabels={00,01,10,11},nodes near coords, enlarge y limits={upper, value=0.22}, ylabel={Count},y label style={at={(axis description cs:0.1,.5)}}]
            \addplot [draw=none,fill=blue] coordinates {
            (1, 37)
            (2, 958)
            (3, 3)
            (4, 26)
            };
        \end{axis}
    \end{tikzpicture}}%
    \subfloat[\centering Measurement results of Bob when Alice encoded `10' as her message bit]{\begin{tikzpicture}
        \tikzstyle{every node}=[font=\small]
        \begin{axis}[ybar,bar width = 10pt, width=0.26\textwidth, xtick={1,2,3,4},xticklabels={00,01,10,11},nodes near coords, enlarge y limits={upper, value=0.22}, ylabel={Count},y label style={at={(axis description cs:0.1,.5)}}]
            \addplot [draw=none,fill=blue] coordinates {
            (1, 15)
            (2, 1)
            (3, 967)
            (4, 41)
            };
        \end{axis}
    \end{tikzpicture}}%
    \subfloat[\centering Measurement results of Bob when Alice encoded `11' as her message bit]{\begin{tikzpicture}
        \tikzstyle{every node}=[font=\small]
        \begin{axis}[ybar,bar width = 10pt, width=0.26\textwidth, xtick={1,2,3,4},xticklabels={00,01,10,11},nodes near coords, enlarge y limits={upper, value=0.22}, ylabel={Count},y label style={at={(axis description cs:0.1,.5)}}]
            \addplot [draw=none,fill=blue] coordinates {
            (1, 3)
            (2, 12)
            (3, 37)
            (4, 972)
            };
        \end{axis}
    \end{tikzpicture}}%
    \caption{Measurement outcomes of Bob, when Alice encodes 2-bit message and sends it using proposed UA-DI-QSDC protocol implemented on $ibm\_brisbane$ with 1024 shots}%
    \label{Fig: Bob Measure}
\end{figure*}

We emulated the UA-DI-QSDC protocol by implementing it on IBM's superconducting qubit-based quantum hardware~\cite{PhysRevLett.91.167005}. We modeled the actions of Alice and Bob within the quantum communication protocol by incorporating both communicating parties' operations into a single quantum circuit framework. To model the non-instantaneous nature of the quantum channels, we represent an ideal quantum channel (without eavesdroppers) as a series of identity gates~\cite{Das2021}. This approach maintains the coherence of quantum information as it is transmitted from Alice to Bob, allowing for a clear depiction of the protocol's execution and the interaction between the quantum states. The use of identity gates ensures that the qubits remain unchanged during the transmission, thereby preserving their quantum state and facilitating an accurate emulation of the communication protocol. However, in practice, these gates are subject to noise, causing the channel to deviate from an ideal identity operation. We analyze the accuracy of message transfer under realistic noisy quantum channels of varying lengths. Further, we simulate four common quantum attacks: impersonation, intercept-and-resend, entangle-and-measure, and man-in-the-middle attacks to assess the protocol's security in real-world conditions.

\subsection{Experimental setup}
The experimental setup primarily utilizes the IBM quantum device $ibm\_brisbane$ for all experiments. This device features $127$ qubits with heavy hexagonal connectivity and is equipped with the $Eagle$ $r3$ processor. The hardware has a $4.5\%$ error per layered gate (EPLG) for a $100$-qubit chain. The device has the median T1 (relaxation time) of $233.04$ $\mu$s, and the median T2 (decoherence time) of $145.75$ $\mu$s. %The device has a median T1: 233.04 $\mu$s and a median T2: 145.75 $\mu$s. 
The median error rate of the identity operator is $2.41\times10^{-4}$ and the execution time of the identity operator is $60 ns$. 

To implement the UA-DI-QSDC protocol in IBM's quantum processors we have used $qiskit 1.1$. The system on which all the experiments have been performed has Python 3.12.4, with processor Intel(R) Core(TM) i5-10310U CPU having a clock speed of 2.21 GHz and 16GB RAM.

\subsection{Performance analysis}
The performance of the UA-DI-QSDC protocol is evaluated based on metrics such as the fidelity of the final measurement outcome compared to the ideal simulation and the error rate in message transmission. The analysis considers different noise levels and qubit coherence times to understand the practical limitations and strengths of the protocol.

\paragraph{Execution on noisy hardware} 
We have executed the protocol in $ibm\_brisbane$ hardware. Each identity gate in this device requires $60 ns$ to execute, and the error probability of each identity gate is $p_e = 2.41\times10^{-4}$. Without loss of generality, we assume $U_{C}=\eta I$, where $\eta \in \mathbb{Z}^+$. Then the probability that the channel remains error-free is $(1 - p_{e})^{\eta}$. However, when executing this circuit, it is subjected to additional sources of error beyond channel noise, such as calibration and readout errors.

In our first experiment, Alice and Bob share one EPR pair. Alice encodes her $2$-bit message on her qubit and sends it to Bob through a quantum channel containing $\eta=10$ identity operators. Alice sends four different messages: `$00$', `$01$', `$10$', and `$11$', through this channel. Bob then measures the EPR pair in the Bell basis and successfully receives the same message that Alice encoded. The measurement outcomes for all four cases are shown in Fig. \ref{Fig: Bob Measure}. The average fidelity of message outcomes is at least 0.95 in all cases.

% \begin{figure*}[!ht]
%     \centering
%     \subfloat[\centering Measurement results of Bob when Alice encoded '00' as her message bit]{{\includegraphics[width=0.21\textwidth]{image (00).png} }}%
%     \qquad
%     \subfloat[\centering  Measurement results of Bob when Alice encoded '01' as her message bit]{{\includegraphics[width=0.21\textwidth]{image (01).png} }}%
%     \qquad
%     \subfloat[\centering Measurement results of Bob when Alice encoded '10' as her message bit]{{\includegraphics[width=0.21\textwidth]{image (10).png} }}%
%     \qquad
%     \subfloat[\centering Measurement results of Bob when Alice encoded '00' as her message bit]{{\includegraphics[width=0.21\textwidth]{image(11).png} }}%
    
%     \caption{Measurement outcomes of Bob, when Alice encodes 2-bit message and sends it using proposed UA-DI-QSDC protocol implemented on $ibm\_brisbane$ with 1024 shots}%
%     \label{Fig: Bob Measure}
% \end{figure*}

\begin{figure}
    \centering
    \includegraphics[width=0.4\textwidth]{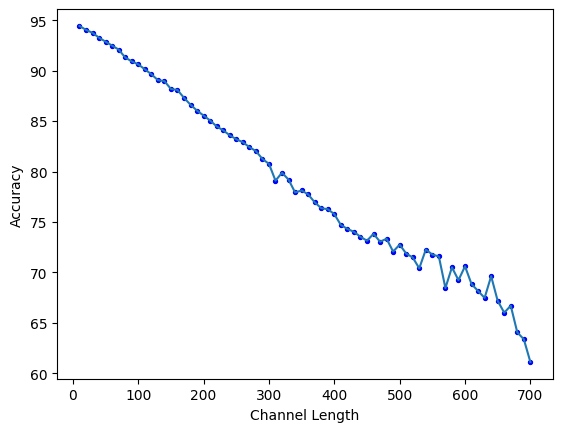}
    \caption{Accuracy of Bob's measurement versus channel length (number of identity operators in the channel)}
    \label{fig: Accuracy vs Channel length}
\end{figure}

\paragraph{Effect of channel length}
We now incorporate quantum channels of varying lengths to analyze the performance of the UA-DI-QSDC protocol. In these subsequent experiments, to simulate the finite duration of the quantum channel, we execute the protocol with $10 \leq \eta \leq 700$ identity gates between Alice and Bob. Since each identity operation in $ibm\_brisbane$ takes $60 ns$ to execute, we have conducted the protocol with the channel of time duration starting from  $0.6\mu s$ to $42 \mu s$. In each experiment, we increased the channel duration by $0.6\mu s$. We have estimated Bob's measurement outcomes and plotted them in Fig.~\ref{fig: Accuracy vs Channel length}.

From Fig.~\ref{fig: Accuracy vs Channel length}, it is evident that beyond 700 identity gates or a duration of $42 \mu$s, the accuracy of the protocol drops below $60\%$. 

To ensure the reliability of the UA-DI-QSDC protocol over longer noisy quantum channels, the use of error-correcting codes~\cite{nielsen00} is essential. These codes can correct errors caused by qubits interacting with the external environment, thereby improving the fidelity of the transmitted quantum states. However, implementing these codes requires a significant number of qubits and gate operations. To avoid usage additional quantum resources, quantum error mitigation or error suppression techniques~\cite{Nautrup_2019,9226505,10.1145/3539613,BASU2024112085} can be adopted. These techniques can help maintain the protocol's reliability without the extensive overhead of error-correcting codes.

\section{Conclusion and future work} \label{conclusion}

%The UA-DI-QSDC protocol represents a significant advancement in the field of quantum communication, offering a secure and authenticated communication framework that is resilient against device imperfections and various attack vectors. The implementation on IBM's quantum computers demonstrates the practical feasibility and robustness of the protocol. 

%Future work will focus on optimizing the protocol for higher communication rates and exploring its integration with existing classical and quantum communication networks. The potential applications of UA-DI-QSDC in secure communications, financial transactions, and critical infrastructure protection highlight its relevance and importance in the evolving landscape of quantum technologies.

The proposed UA-DI-QSDC protocol marks a substantial advancement in the field of quantum communication by providing a secure and authenticated communication framework that is resilient against both device imperfections and a variety of attacks. The successful implementation of the protocol on IBM's quantum hardware further demonstrate its practical feasibility and robustness, confirming its potential for real-world application.

Looking ahead, our future research will focus on optimizing the protocol to achieve higher communication rates, as well as exploring its seamless integration with existing classical and quantum communication networks. The broad range of potential applications for the UA-DI-QSDC protocol, including secure communications, financial transactions, and the protection of critical infrastructure, underscores its significant relevance and importance in the rapidly evolving landscape of quantum technologies.
\bibliographystyle{IEEEtran}
\bibliography{main}

\end{document}